\documentclass[twocolumn,aps,pre,showpacs,preprintnumbers]{revtex4}
\usepackage{amssymb,graphicx}

\begin{document}
\title{Concise theory of chiral lipid membranes}

\author{Z. C. Tu}
\affiliation{\textrm{II.} Institut f\"{u}r Theoretische Physik,
Universit\"{a}t Stuttgart, Pfaffenwaldring 57, 70550 Stuttgart,
Germany}
\author{U. Seifert}
\affiliation{\textrm{II.} Institut f\"{u}r Theoretische Physik,
Universit\"{a}t Stuttgart, Pfaffenwaldring 57, 70550 Stuttgart,
Germany}

\begin{abstract} A theory of chiral lipid membranes is proposed on the basis of
a concise free energy density which includes the contributions of
the bending and the surface tension of membranes, as well as the
chirality and orientational variation of tilting molecules. This
theory is consistent with the previous experiments [J.M. Schnur
\textit{et al.}, Science \textbf{264}, 945 (1994); M.S. Spector
\textit{et al.}, Langmuir \textbf{14}, 3493 (1998); Y. Zhao,
\textit{et al.}, Proc. Natl. Acad. Sci. USA \textbf{102}, 7438
(2005)] on self-assembled chiral lipid membranes of DC$_{8,9}$PC. A
torus with the ratio between its two generated radii larger than
$\sqrt{2}$ is predicted from the Euler-Lagrange equations. It is
found that tubules with helically modulated tilting state are not
admitted by the Euler-Lagrange equations, and that they are less
energetically favorable than helical ripples in tubules. The pitch
angles of helical ripples are theoretically estimated to be about
0$^\circ$ and 35$^\circ$, which are close to the most frequent
values 5$^\circ$ and 28$^\circ$ observed in the experiment [N.
Mahajan \textit{et al.}, Langmuir \textbf{22}, 1973 (2006)].
Additionally, the present theory can explain twisted ribbons of
achiral cationic amphiphiles interacting with chiral tartrate
counterions. The ratio between the width and pitch of twisted
ribbons is predicted to be proportional to the relative
concentration difference of left- and right-handed enantiomers in
the low relative concentration difference region, which is in good
agreement with the experiment [R. Oda \textit{et al.}, Nature
(London) \textbf{399}, 566 (1999)].

\pacs{68.15.+e, 87.10.+e, 61.30.-v} \preprint{Phys. Rev. E
\textbf{76}, 031603 (2007)}
\end{abstract}
\maketitle

\section{Introduction}
Since tubules were fabricated successfully from chiral lipid
molecules, chiral lipid structures have attracted much experimental
attention
\cite{Schnur93,Schnur94,Spector98,Zhaoy05,Zhaoy06,Spector01,Chung93,Zastavker,Oda99,Oda02}.
Most of chiral structures in experiments are self-assembled from
DC$_{8,9}PC$ which is a typical chiral molecule. Schnur \emph{et
al.} have observed that the spherical vesicles in solution have very
weak circular dichroism signal while tubules have strong one
\cite{Schnur94}. Spector \emph{et al.} have investigated the chiral
lipid tubules formed from various proportions of left- and
right-handed DC$_{8,9}PC$ molecules and found that they are of
similar radii \cite{Spector98}, which reveals that the radii of
chiral tubules do not depend on the strength of the molecular
chirality. Fang's group has carefully resolved the molecular tilting
order in the tubules and concluded that the projected direction of
the molecules on the tubular surfaces departs 45$^\circ$ from the
equator of the tubules at the uniform tilting state \cite{Zhaoy05}.
Helical ripples in lipid tubules are also observed by the same group
with atomic force microscopy \cite{Zhaoy06}. Their pitch angles are
found to be concentrated on about 5$^\circ$ and 28$^\circ$
\cite{Zhaoy06}. Cholesterol is another kind of chiral molecules used
in the experiments where helical stripes with pitch angles
$11^\circ$ and $54^\circ$ are usually observed
\cite{Chung93,Zastavker}. Additionally, Oda \emph{et al.} have
reported twisted ribbons of achiral cationic amphiphiles interacting
with chiral tartrate counterions \cite{Oda99,Oda02}. It is found
that the twisted ribbons can be tuned by the introduction of
opposite-handed chiral counterions in various proportions
\cite{Oda99}. From the experimental data \cite{Oda99}, we see that
the ratio between the width and pitch of the ribbons is proportional
to the relative concentration difference of left- and right-handed
enantiomers in the low relative concentration difference region. Can
we interpret all or at least most of above experimental results
within a unified theory?

There are several theoretical discussions on chiral lipid membranes
(CLMs) in the previous literature, where the chiral molecules are
assumed to be in a Smectic C$^*$ phase at which the direction of the
molecules is tilted from the normal of the membranes at a constant
angle. The possible free energy of CLMs is discussed by Helfrich and
Prost from symmetry arguments \cite{Helfrich88}. Their theory has
been further developed and applied in many studies
\cite{oy90,Nelson92,Selinger93,Selinger96,Selinger01,oy98,HuLo05}.
Nelson and Powers have investigated the thermal fluctuations of CLMs
\cite{Nelson92}. Selinger \emph{et al.} have discussed tubules with
helically modulated tilting state and helical ripples in tubules
\cite{Selinger93,Selinger96,Selinger01}. Komura and Ou-Yang have
given an explanation to the high-pitch helical stripes of
cholesterol molecules \cite{oy98}. Due to the complicated form of
the free energy used in these theories
\cite{Helfrich88,oy90,Nelson92,Selinger93,Selinger96,Selinger01,oy98},
it is almost impossible to obtain the general Euler-Lagrange
equations corresponding to the free energy. Thus one cannot
determine whether a configuration, such as a twisted ribbon, a
tubule with helically modulated tilting state or a helical stripe,
is a true equilibrium structure or not. This difficulty was always
ignored in previous discussion of both tubules with helically
modulated tilting state and helical stripes
\cite{Selinger93,Selinger96,Selinger01,oy98}. Can we confront this
difficulty and construct a more concise theory of CLMs consistent
with the experiments, in which we can unambiguously say which
configuration is a genuine equilibrium structure?

We will address these questions in this paper, which is organized as
follows: In Sec.~\ref{Endensity}, we introduce a concise free energy
density of CLMs which includes the contributions of the bending and
the surface tension of the membranes, as well as the chirality and
orientational variation of the tilting molecules. In
Sec.~\ref{ClosedCLMs}, we present the Euler-Lagrange equations for
CLMs without free edges and use them to explain experimental data
\cite{Schnur94,Spector98,Zhaoy05,Zhaoy06}. We predict a torus with
the ratio between its two generated radii larger than $\sqrt{2}$,
which has not yet been observed in the experiments on self-assembled
CLMs. In Sec.~\ref{openCLMS}, we present the Euler-Lagrange
equations and boundary conditions for CLMs with free edges and use
them to discuss experimental data \cite{Chung93,Zastavker,Oda99}.
Sec.~\ref{summary} is a brief summary and discussion. In the
Appendixes, we briefly derive the Euler-Lagrange equations for CLMs
without free edges and the Euler-Lagrange equations as well as
boundary conditions for CLMs with free edges through the variational
method \cite{tzcjpa} developed by one of the present authors.
Several mathematical details are also put in the Appendixes.

\section{Free energy density \label{Endensity}}
Following the above theories
\cite{Helfrich88,oy90,Nelson92,Selinger93,Selinger96,Selinger01,oy98},
we adopt a concise form of free energy density for a CLM which
consists of the following contributions.

(i) The bending and surface energy per area is taken as Helfrich's
form \cite{helfrich}
\begin{equation}G_{H}=(k_{c}/2)(2H+c_{0})^{2}-\bar{k}K+\lambda,\label{frhelfr}\end{equation}
where $k_c$ and $\bar{k}$ are bending rigidities, and $\lambda$ the
surface tension. $c_0$ is the spontaneous curvature reflecting the
asymmetrical factors between two sides of the membrane. $H$ and $K$
are the mean curvature and Gaussian curvature of the membrane,
respectively, which can be expressed as $2H=-(1/R_1+1/R_2)$,
$K=1/R_1R_2$ by the two principal curvature radii $R_1$ and $R_2$.
The curvature energy in Eq.\,(\ref{frhelfr}) is invariant under the
coordinate rotation around the normal of the membrane surface, but
will change under the inversion of the normal if $c_0\neq 0$.

\begin{figure}[pth!]
\includegraphics[width=7cm]{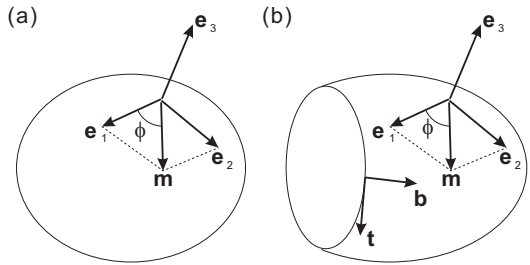}\caption{\label{figframe}
Right-handed orthonormal frame
$\{\mathbf{e}_1,\mathbf{e}_2,\mathbf{e}_3\}$ at any point in a
surface where $\mathbf{e}_3$ is the normal vector of the surface.
(A) Surface without boundary curve. (B) Surface with boundary curve
where $\mathbf{t}$ is the tangent vector of the boundary curve, and
$\mathbf{b}$, in the tangent plane of the surface, is perpendicular
to $\mathbf{t}$.}
\end{figure}

(ii) The energy per area originating from the chirality of tilting
molecules has the form \cite{oy90}
\begin{equation}G_{ch}=-h\tau_{\mathbf{m}},\label{chiralen}\end{equation}
where $h$ reflects the strength of the molecular chirality which
usually determines the handedness of CLMs---the CLMs with the
opposite handedness will be observed in experiments if $h$ changes
its sign \cite{Schnur94}. Without losing the generality, we only
discuss the case of $h>0$ in this paper. $\tau_{\mathbf{m}}$ is the
geodesic torsion along the unit vector $\mathbf{m}$ at some point.
Here $\mathbf{m}$ represents the projected direction of the lipid
molecules in the experiments
\cite{Schnur94,Spector98,Zhaoy05,Zhaoy06,Spector01,Chung93,Zastavker}
and chiral tartrate counterions in the experiment \cite{Oda99,Oda02}
on the membrane surface, respectively. If we take a right-handed
orthonormal frame $\{\mathbf{e}_1,\mathbf{e}_2,\mathbf{e}_3\}$ with
$\mathbf{e}_3$ being the normal vector of the membrane as shown in
Fig.~\ref{figframe}, $\mathbf{m}$ can be expressed as
$\mathbf{m}=\cos\phi\mathbf{e}_1+\sin\phi\mathbf{e}_2$, where $\phi$
is the angle between $\mathbf{m}$ and $\mathbf{e}_1$. At this frame,
the curvature tensor can be expressed as a matrix with element $a$,
$b$, and $c$ shown in Appendix~\ref{Appdmfm}. With the curvature
tensor, the geodesic torsion along $\mathbf{m}$ can be expressed as
\cite{tzcjpa}
\begin{equation}\tau _{\mathbf{m}} =b(\cos^2\phi-\sin^2\phi)+(c-a)\cos\phi\sin\phi,\label{torsionexp}\end{equation}
which is the same as the chiral term in the previous literature,
such as the last term of Eq.\,(2) in Ref.~\cite{Helfrich88},
Eq.\,(3) in Ref.~\cite{Nelson92}, and the third term of Eq.\,(2.1)
in Ref.~\cite{Selinger96}. In particular, for the principal frame,
the above equation is simplified as
\begin{equation}\tau_{\mathbf{m}}=(1/R_1-1/R_2)\cos\phi\sin\phi.\label{torsioninpr}\end{equation}
We will confine $\phi$ to the region ($-\pi/2,\pi/2$] because of the
relation $\tau_{\mathbf{m}}(\phi+\pi)=\tau_{\mathbf{m}}(\phi)$.
Moreover, it is easy to see that the geodesic torsion along the
mirror image of $\mathbf{m}$ with respect to $\mathbf{e}_1$ changes
its sign because $\phi\mapsto -\phi$ under the reflection with
respect to $\mathbf{e}_1$. Thus this term breaks the inversion
symmetry in the tangent plane at each point of the membrane, which
allows us to distinguish the handedness. Additionally, from
Eq.\,(\ref{torsioninpr}) we can see that the minimum of
Eq.\,(\ref{chiralen}),
\begin{equation}G_{ch}^{min}=-(h/2)|1/R_1-1/R_2|,\end{equation} is reached when $\mathbf{m}$
departs from the principal direction at a angle $+\pi/4$ or $-\pi/4$
for a given shape of the membrane. Here the sign in front of $\pi/4$
depends on the sign of $(1/R_1-1/R_2)$. The larger difference
between $R_1$ and $R_2$ is, the larger absolute value of
$G_{ch}^{min}$ is reached. In other words, the chiral term favors
saddle surfaces (for example, the twisted ribbons in
Sec.~\ref{SecTwR}) whose two principal curvature radii $R_1$ and
$R_2$ have opposite signs.

(iii) The energy per area due to the orientational variation of
$\mathbf{m}$ is taken as \cite{Lubensky92}
\begin{equation}G_{ov}=(k_{f}/2)[(\nabla\times\mathbf{m})^{2}+(\nabla\cdot\mathbf{m})^{2}],\end{equation}
where $k_f$ is a constant in the dimension of energy. This is the
simplest term of energy cost due to tilting order invariant under
the coordinate rotation around the normal of the membrane surface.
$\nabla$ is the 2-dimensional (2D) gradient operator on the membrane
surface, and the 2D cross product ``$\times$'' gives a scalar. By
defining a spin connection field $\mathbf{S}$ satisfying
$\nabla\times\mathbf{S}=K$ \cite{Nelson87}, one can derive
\begin{equation}(\nabla\times\mathbf{m}) ^{2}+(
\nabla\cdot\mathbf{m})^{2}=(\nabla\phi-\mathbf{S})^{2}\end{equation}
through simple calculations.

The total free energy density adopted in the present paper,
$G=G_{H}+G_{ch}+G_{ov}$, has the following concise form:
\begin{equation}
G=\frac{k_{c}}{2}(2H+c_{0})^{2}-\bar{k}K+\lambda-h\tau_{\mathbf{m}}+\frac{k_{f}}{2}\mathbf{v}
^{2},\label{energy2}
\end{equation}
where $\mathbf{v}\equiv\nabla\phi-\mathbf{S}$. This special form
might arguably be the most natural and concise construction
including the bending, chirality and tilting order, for the given
vector field $\mathbf{m}$ and normal vector field $\mathbf{e}_3$. Of
course, using these two fields, one can also construct more general
free energy densities which contain much more terms
\cite{Helfrich88,oy90,Nelson92,Selinger93,Selinger96,Selinger01,oy98}.
Compared with the general form, our special form is simplified in
two aspects: (i) The contributions of the bending of membranes and
the orientational elasticity of $\mathbf{m}$ are taken as the
isotropic forms; (ii) There is no additional coupling between
$\mathbf{m}$ and the curvature of membranes except the chiral term
(\ref{chiralen}), which makes it a concise theory. Although our free
energy density is not new relative to the general form appearing in
the previous literature, we derive for the first time the
corresponding Euler-Lagrange equations without any assumption on the
shapes of membranes before doing the variation, and then check which
configuration solves the Euler-Lagrange equations. More importantly,
in the following, we will see that most of the experimental results
can be explained in terms of such a concise free energy density.

\section{Chiral lipid membranes without free edges \label{ClosedCLMs}}
CLMs without free edges usually correspond to closed vesicles. Here
a long enough tubule, where the end effect is neglected, is also
regarded as a CLM without free edges. In this section, we show the
Euler-Lagrange equations of CLMs without free edges (derived briefly
in Appendix~\ref{appendelclose}), and then discuss CLMs in
spherical, tubular and torus shapes. As we will see, our theoretical
results are in good agreement with the experiments
\cite{Schnur94,Spector98,Zhaoy05,Zhaoy06}.

\subsection{Euler-Lagrange equations}
The free energy for a closed CLM can be expressed as
\begin{equation}F=\int G\ dA +P\int dV,\label{closedFE}\end{equation}
where $dA$ is the area element of the membrane and $dV$ the volume
element enclosed by the vesicle. $P$ is either the pressure
difference between the outer and inner sides of the vesicle or used
to implement a volume constraint. For tubular configuration, we
usually take $P=0$.

Using the variational method \cite{tzcjpa}, we obtain the
Euler-Lagrange equations corresponding to the free energy
(\ref{closedFE}) as
\begin{equation}
2h(\kappa_{\mathbf{m}}-H)-k_{f}\nabla^{2}\phi=0, \label{EL1}
\end{equation}
and
\begin{eqnarray}
&&\hspace{-0.5cm}2k_{c}\nabla^{2}H+k_{c}( 2H+c_{0}) (
2H^{2}-c_{0}H-2K)-2\lambda H+P\nonumber\\
&&\hspace{-0.5cm}+h[  \nabla\cdot( \mathbf{m}\nabla\times\mathbf{m})
+\nabla\times(
\mathbf{m}\nabla\cdot\mathbf{m})]\nonumber\\
&&\hspace{-0.5cm}+k_{f}[( \kappa_{\mathbf{v}}-H) \mathbf{v}^{2}-
\nabla\mathbf{v}\colon\nabla\mathbf{e}_3] =0,\label{EL2}
\end{eqnarray}
where $\kappa_{\mathbf{m}}$ and $\kappa_{\mathbf{v}}$ are the normal
curvature along the directions of $\mathbf{m}$ and $\mathbf{v}$,
respectively. They can be expressed as Eqs.\,(\ref{normalcur}) and
(\ref{norcurinv}) with the curvature tensor. Physically,
Eqs.\,(\ref{EL1}) and (\ref{EL2}) express the moment and force
balances along the normal $\mathbf{e}_3$ at each point of the
membrane. If $h$ and $k_f$ vanish, the above two equations
degenerate into the shape equation of achiral lipid vesicles, which
has been fully discussed in
Refs.~\cite{oy87,Lipowsky91,Seifert97,oybook,DuLiuWang06}.

Two remarks are necessary concerning Eq.\,(\ref{EL1}): (i) We have
selected the proper gauge such that $\nabla\cdot\mathbf{S}=0$, or
else $\nabla^{2}\phi$ should be replaced with
$\nabla^{2}\phi-\nabla\cdot\mathbf{S}$. (ii) For closed vesicles
different from toroidal topology, the tangent vector field
$\mathbf{m}$ will have singular points. In this case, the right-hand
term should be replaced with the sum of $\delta$-function,
$\sum_i\sigma_i\delta(\mathbf{r}-\mathbf{r}_i)$, where $\mathbf{r}$
and $\mathbf{r}_i$ represent any point and singular point in the
CLM, while $\sigma_i$ represents the strength of the source or
vortex at the singular point $\mathbf{r}_i$.

\subsection{Spherical vesicles}
For spherical vesicles of chiral lipid molecules,
$\tau_{\mathbf{m}}$ is always vanishing because $R_1=R_2$. Thus the
free energy (\ref{closedFE}) is independent of the molecular
chirality and permits the same probability of left- and right-handed
spherical vesicles existing in solution. Naturally, no evident
circular dichroism signal would be observed, which is consistent
with the experiment \cite{Schnur94}. Of course, we cannot exclude
the other possible explanation of this experiment that the lipid
molecules are non-tilting in the spherical vesicles
\cite{Selinger01}.

\subsection{Tubules with uniform tilting state}
\begin{figure}[pth!]
\includegraphics[width=7cm]{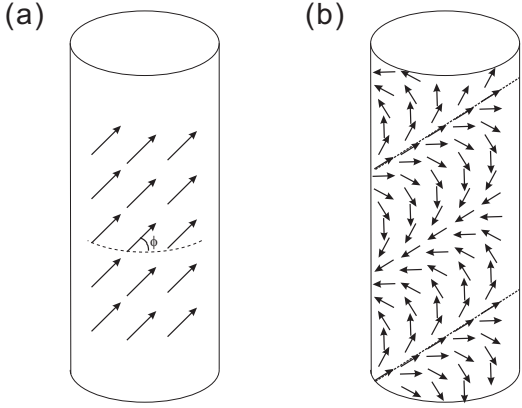}\caption{\label{tubules}
Tubules: (a) Uniform tilting state; (b) Helically modulated tilting
state. Arrows represent the projected directions $\{\mathbf{m}\}$ of
the tilting molecules on the tubules.}
\end{figure}

A tubule is regarded as an cylinder with radius $\rho$ and infinite
length. Then $H=-1/2\rho$, $\kappa_{\mathbf{m}}=-\cos^2\phi/\rho$,
and $\tau_{\mathbf{m}}=\sin2\phi/2\rho$, where $\phi$ is the angle
between $\mathbf{m}$ and the equator of the cylinder. $\phi$ is a
constant for the tubule with the uniform tilting state as shown in
Fig.~\ref{tubules}(a). Thus Eq.\,(\ref{EL1}) requires
\begin{equation}\phi=\pm\pi/4.\end{equation}
We should keep $\phi=\pi/4$ because it corresponds to the local
minimum of $-h\int\tau_{\mathbf{m}}dA$, which is in good agreement
with the experiment by Fang's group \cite{Zhaoy05} where the
projected direction of the molecules on the tubular surfaces indeed
departing 45$^\circ$ from the equator of the tubules with the
uniform tilting state.

For the cylindrical shape, we set $P=0$. It is easy to find
$\nabla\cdot( \mathbf{m}\nabla\times\mathbf{m}) +\nabla\times(
\mathbf{m}\nabla\cdot\mathbf{m})$ and $( \kappa_{\mathbf{v}}-H)
\mathbf{v}^{2}- \nabla\mathbf{v}\colon\nabla\mathbf{e}_3$ both to be
vanishing for $\phi=\pi/4$. Thus Eq.\,(\ref{EL2}) is transformed
into
\begin{equation}
k_c(c_0^2\rho^2-1)+2\lambda\rho^2=0.\label{tubradius}
\end{equation}
This equation indicates that the radius $\rho$ of chiral lipid
tubules is independent of the strength of molecular chirality $h$,
which is consistent with the experiment \cite{Spector98} where the
tubular radius is insensitive to the proportion of left- and
right-handed DC$_{8,9}PC$ in the tubule. However, we cannot exclude
the other possible explanation that DC$_{8,9}PC$ molecules with
different handedness are not mixed in the tubule such that each
tubule in solution contains merely the one kind of molecules with
the same handedness ether left- or right-handed
\cite{Spector98,Selinger01}.

\subsection{Tubules with helically modulated tilting state\label{SecHMTS}}
The concept of tubules with helically modulated tilting state was
proposed by Selinger \emph{et al.}~\cite{Selinger96}. The
orientational variation of the tilting molecules is assumed to be
linear and confined to a small range. Here we will directly
investigate the orientational variation by using the Euler-Lagrange
equation (\ref{EL1}) without this assumption. Additionally, as
mentioned in the Introduction, whether the tubules with helically
modulated titling state are equilibrium configurations or not is not
addressed in Ref.~\cite{Selinger96}. However, our theory will
unambiguously reveal that they are not equilibrium configurations.
Let $s$ and $z$ denote the arc length parameters along the
circumferential and axial directions, respectively. If $\phi$, the
angle between $\mathbf{m}$ and the equator of the cylinder, is not a
constant, Eqs.\,(\ref{EL1}) and (\ref{EL2}) are then transformed
into
\begin{equation}k_{f}(\phi_{ss}+\phi_{zz})+(h/\rho)\cos2\phi=0,\label{stripe01}\end{equation}
and
\begin{eqnarray}&&\hspace{-0.4cm}h[ 2(
\phi_{z}^{2}-\phi_{s}^{2}+\phi_{sz})\sin2\phi+(
\phi_{ss}-\phi_{zz}+4\phi_{z}\phi_{s})\cos2\phi]\nonumber\\&&\hspace{-0.4cm}+\lambda/\rho+k_{c}(
c_{0}^{2}-1/\rho^{2}) /2\rho+k_{f}[ (
\phi_{z}^{2}-\phi_{s}^{2})/2\rho+\phi_{sz}/\rho]\nonumber\\&&\hspace{-0.4cm}=0.\label{stripe02}\end{eqnarray}
where the subscripts $s$ and $z$ represent the partial derivatives
respect to $s$ and $z$, respectively. The derivations of the above
two equations are shown in Appendix~\ref{deriveqshelic}.

At the helically modulated tilting state, $\phi$ is invariant along
the direction of a fictitious helix enwinding around the tubule as
shown in Fig.~\ref{tubules}(b). Let $\psi$ be the pitch angle of
that helix and apply a coordinate transformation
$(s,z)\rightarrow(\zeta,\eta)$ via $\zeta=s\cos\psi+z\sin\psi, \eta
=-s\sin\psi+z\cos\psi$, where $\zeta$ is the coordinate along the
helix and $\eta$ is the coordinate orthogonal to $\zeta$. In the new
coordinates, $\phi$ depends only on $\eta$. Changing variable
$\Theta=\phi-\psi$ and introducing the dimensionless parameters
$\chi=\eta/\rho$ and $\bar{h}=h\rho/k_f$, we transform
Eqs.\,(\ref{EL1}) and (\ref{EL2}), respectively, into
\begin{equation}\Theta_{\chi\chi}=-\bar{h}\cos2(\Theta+\psi),\label{stripe1}\end{equation}
and
\begin{equation}(k_c/k_f)(1-c_0^2\rho^2)-2\lambda\rho^2/k_f=R,\label{stripe2}
\end{equation}
where $R\equiv\Theta_{\chi}^{2}\cos2\psi-\Theta_{\chi\chi}\sin2\psi
+2\bar{h}( 2\Theta_{\chi}^{2}\sin2  \Theta -\Theta_{\chi\chi}\cos2
\Theta)$.

The first integral of Eq.\,(\ref{stripe1}) is
\begin{equation}\Theta_{\chi}^{2}=\mu^{2}-\bar{h}\sin2(\Theta+\psi), \label{integral}\end{equation}
with an unknown constant $\mu$. Substituting Eqs.\,(\ref{stripe1})
and (\ref{integral}) into Eq.\,(\ref{stripe2}), we obtain the
right-hand side of Eq.\,(\ref{stripe2}):
\begin{eqnarray}R&=&\mu^{2}\cos2\psi+\bar{h}(4\mu^{2}-1)\sin2\Theta\nonumber\\&+&2\bar{h}^2[\cos(4\Theta+2\psi)-\sin2\Theta\sin2(\Theta+\psi)].\label{rhtR}\end{eqnarray}
The necessary condition for validity of Eq.\,(\ref{stripe2}) is that
the right-hand side $R$ is constant, which, in terms of
Eq.\,(\ref{rhtR}), holds if and only if $\bar{h}=0$ for varying
$\Theta$. Therefore, tubules with helically modulated tilting state
are not admitted by the Euler-Lagrange equation (\ref{EL2}) if
$h\neq 0$. In other words, the orientational variation of the
tilting lipid molecules breaks the force balance along the normal
direction of the tubule at this state, which might rather induce
helical ripples in tubules.

\subsection{Helical ripples in tubules\label{Secripple}}
Assume now that a tubule with radius $\rho$ undergoes small
out-of-plane deformations and reaches a new configuration expressed
as a vector $\{\rho( 1+y) \cos( s/\rho),\rho(1+y) \sin(s/\rho)
,z\}$, where $|y|\ll 1$ is a function of $s$ and $z$. For
simplicity, we take $c_0=0$, $\lambda=0$, $k_f\simeq k_c$, and
$\bar{h}\ll 1$ in this subsection. As shown in
Appendix~\ref{dereqsripp}, Eqs.\,(\ref{EL1}) and (\ref{EL2}) can be
transformed into
\begin{equation}\rho^2(\phi_{ss}+\phi_{zz}+\phi_{z}y_{z}-\phi_{s}y_{s}-2y\phi
_{ss}-y_{zs})=-\bar{h}\cos2\phi\label{eqripp1}\end{equation} and
\begin{eqnarray}
&&\hspace{-0.15cm}(k_c/k_f\rho^2)[(1+\rho^{2}\partial_{ss}+\rho^{2}
\partial_{zz})^{2}y-1/2]\nonumber\\
&&\hspace{-0.4cm}+(k_c/k_f\rho^2)[(1+\rho^{2}\partial_{ss}+\rho^{2}
\partial_{zz})y/2-2\rho^{2}y_{zz}]\nonumber\\
&&\hspace{-0.4cm}+\bar{h}[2(\phi_z^2-\phi_s^2+\phi_{sz})\sin2\phi
+(\phi_{ss}-\phi_{zz}+4\phi_{z}\phi_{s})\cos2\phi]\nonumber\\
&&\hspace{-0.4cm}+(\phi_s^2-\phi_z^2)(y+\rho^{2}y_{ss}-\rho^{2}y_{zz}-1)/2\nonumber\\
&&\hspace{-0.4cm}+\phi_{sz}+y\phi_{s}^{2}-2y\phi_{sz}-y_{zz}+2\rho^2
y_{sz}\phi_{s}\phi_{z}=0,\label{eqripp2}
\end{eqnarray} where $\phi$ is the
angle between $\mathbf{m}$ and the equator of the tubule.

\begin{figure}[pth!]
\includegraphics[height=6cm]{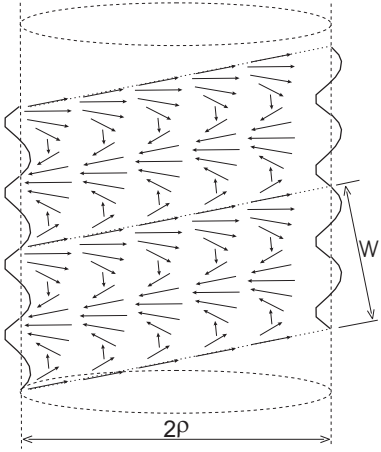}\caption{\label{ripplesfig}
Small amplitude ripples in a tubule with radius $\rho$. Arrows
represent the projected directions $\{\mathbf{m}\}$ of the tilting
molecules on the ripples' surface.}
\end{figure}

Now we consider helical ripples in the tubule where $\phi$ and $y$
are invariant along the direction of a fictitious helix enwinding
around the tubule as shown in Fig.~\ref{ripplesfig}. We adopt the
same coordinate transformation as the above subsection, and let
$\vartheta=\phi-(\Theta+\psi)$, where $\psi$ is the pitch angle of
the fictitious helix and $\Theta$ is governed by
Eq.\,(\ref{stripe1}). Then the above equations (\ref{eqripp1}) and
(\ref{eqripp2}) are reduced to a matrix equation
\begin{equation}\mathcal{L}\Psi=\Phi,\label{rippl2}\end{equation}
with $\Psi\equiv\{\vartheta,y\}^{t}$ and $\Phi\equiv\{0,(\bar{h}/2)(
1-4\mu^{2})\sin2\Theta\}^t$, where the superscript `$t$' represents
the transpose. The differential operator $\mathcal{L}$ has four
matrix elements:
\begin{eqnarray*}
&&\mathcal{L}_{11}\equiv-{d^{2}}/{d\chi^{2}},\\
&&\mathcal{L}_{12}
\equiv-\mu\cos2\psi{d}/{d\chi}-\sin\psi\cos\psi{d^{2}
}/{d\chi^{2}},\\
&&\mathcal{L}_{21}\equiv\mu\cos2\psi{d}/{d\chi}-\sin\psi\cos\psi
{d^{2}}/{d\chi^{2}},\\
&&\mathcal{L}_{22}
\equiv\mu^{2}\cos^{2}\psi+\mu^{2}\cos2\psi{d^{4}}/{d\chi^{4}}\\&&\hspace{6.5mm}+[
2\mu^{2}\cos2\psi\sin ^{2}\psi+( \mu^{2}-1)  \cos^{2}\psi] {d^{2}}/
{d\chi^{2}},
\end{eqnarray*}
where $\mu$ satisfies
\begin{equation}
\mu^{2}\cos2\psi=k_c/k_f.\label{rippl1}\\
\end{equation}

It is not hard to find a special solution of Eq.\,(\ref{rippl2}) as
\begin{equation}\tilde{\Psi}=\frac{\bar{h}(
1-4\mu^{2})}{4\Gamma}\left(\begin{array}{l}\cos2(\Theta+\psi)\\
2\sin2\Theta\end{array}\right)\label{ripplsolution}\end{equation}
with $\Gamma\equiv\mu^{2}(3\cos^{2}\psi+\cos2\psi-8\mu^{2}\cos2\psi\sin^{2}%
\psi-4\mu^{2}\cos^{2}\psi+16\mu^{4}\cos2\psi)$. Considering
Eq.\,(\ref{rippl1}), we can prove $\Gamma>0$ for $k_c/k_f>0.133$ and
hence $\Gamma>0$ for $k_c\simeq k_f$.

Note that Eq.\,(\ref{rippl2}) contains only the first order terms of
$\vartheta$ and $y$, which can also be obtained from the variation
of the free energy $F$ expanded up to the second order terms of
$\vartheta$ and $y$. The dimensionless mean energy difference
between the tubule with helical ripples and that with helically
modulated tilting state is expressed as
\begin{equation}\Delta F=\int_{0}^{W/\rho}\left(  \frac{1}{2}\Psi
^{t}\mathcal{L}\Psi-\Psi^{t}\Phi\right)  d\chi,
\end{equation} where $W$ is the period of the ripples along
$\eta$ direction as show in Fig.~\ref{ripplesfig}. Substituting the
solution (\ref{ripplsolution}) into the above equation, we have
\begin{equation}
\Delta F=-\frac{\bar{h}^2(1- 4\mu ^{2})^{2}}{8
\Gamma}\int_{0}^{W/\rho}\sin^{2}2 \Theta d\chi<0,
\end{equation}
which reveals that a tubule with ripples is energetically more
favorable than that in a helically modulated titling state.
Additionally, we can easily prove from Eq.\,(\ref{rippl1}) that the
pitch angle obeys $\psi<45^\circ$. All ripples in tubules resolved
in the recent experiment \cite{Zhaoy06} with atomic force microscope
have indeed the pitch angles smaller than $45^\circ$. The tubules
with `helically modulated tilting state' observed in the experiment
\cite{Lvov} might also be the tubules with helical ripples whose
amplitudes are below the experimental resolution. In this
experiment, all pitch angles are also smaller than $45^\circ$.

Considering $\bar{h}\ll 1$, we have $\Theta_\chi\approx \mu$ from
Eq.\,(\ref{integral}) and then
\begin{equation}\mu W/\rho\approx 2\pi,\label{matchcond}\end{equation}
due to $W$ being the period of the ripples along $\eta$ direction.
If the pitch angle $\psi=0$, the fictitious helix is a circle, and
Eqs.\,(\ref{rippl1}) and (\ref{matchcond}) requires
$2\pi\rho/W\approx \sqrt{k_c/k_f}$. If $\psi\neq 0$, the fictitious
helix is indeed a helix satisfying $W=2\pi\rho\sin\psi$. It is then
easy to derive
\begin{equation}\cos2\psi/\sin^2\psi\approx k_c/k_f\label{pitchangle}\end{equation}
from Eqs.\,(\ref{rippl1}) and (\ref{matchcond}). Therefore, two
kinds of ripples are permitted in our theory: one has pitch angle
about $0^\circ$; another satisfies Eq.\,(\ref{pitchangle}), which
gives $\psi\simeq 35^\circ$ for $k_c\simeq k_f$. Our theoretical
results are thus close to the most frequent pitch angles (about
5$^\circ$ and 28$^\circ$) of the ripples in tubules observed by
Fang's group \cite{Zhaoy06}.

\begin{figure}[pth!]
\includegraphics[width=5.5cm]{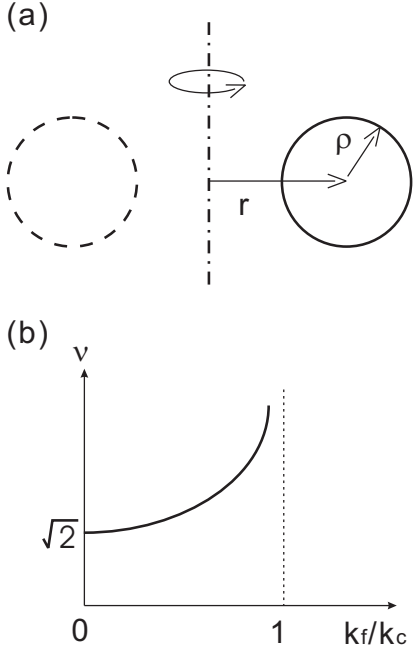}\caption{\label{torusfig}
(a) Torus generated by a cycle rotating around an axis in the same
plane of the cycle. (b) Ratio of the generated radii ($\nu$),
Eq.\,(\ref{torusrad}), as a function of $k_c/k_f$.}
\end{figure}

\subsection{Torus, a prediction\label{Sectorus}}
A torus is a revolution surface generated by a cycle with radius
$\rho$ rotating around an axis in the same plane of the cycle as
shown in Fig.~\ref{torusfig}(a). The revolution radius $r$ should be
larger than $\rho$. A point in the torus can be expressed as a
vector
$\{(r+\rho\cos\varphi)\cos\theta,(r+\rho\cos\varphi)\sin\theta,\rho\sin\varphi\}$.
As shown in Appendix~\ref{dereqstorus}, Eq.\,(\ref{EL1}) is
transformed into
\begin{equation}\frac{\phi_{\theta\theta}}{\nu+\cos\varphi}+[(\nu+\cos\varphi)  \phi_{\varphi}]_\varphi-\frac{\nu h\rho}{k_{f}}\cos2\phi=0,\label{phitorus}\end{equation}
where $\phi$ is the angle between $\mathbf{m}$ and the latitude of
the torus, while $$\nu\equiv r/\rho$$ is the ratio between two
generated radii of the torus.

It is found that the uniform tilting state $\phi=-\pi/4$ satisfies
the above equation (\ref{phitorus}) and that makes $-\int
h\tau_\mathbf{m}dA$ to take the minimum by considering
Eq.\,(\ref{tautours}). As shown in Appendix~\ref{dereqstorus}, with
$\phi=-\pi/4$, Eq.\,(\ref{EL2}) is transformed into
\begin{eqnarray}
&&\hspace{-0.5cm}(2k_c-k_f)/{\nu^2}+k_c(c_0^2\rho^2-1)+2(P\rho+\lambda)\rho^2\nonumber\\
&&\hspace{-0.5cm}+\frac{4 k_c c_0^2\rho^2-4 k_c c_0\rho-2 h \rho+8\lambda \rho^2+6P\rho^3}{\nu}\cos\varphi\nonumber\\
&&\hspace{-0.5cm}+\frac{5 k_c c_0^2\rho^2-8 k_c c_0\rho-4 h \rho+10\lambda \rho^2+3 k_f+6P\rho^3}{\nu^2}\cos^2\varphi\nonumber\\
&&\hspace{-0.5cm}+\frac{2 k_c c_0^2\rho^2-4 k_c c_0\rho-2 h
\rho+4\lambda \rho^2+2
k_f+2P\rho^3}{\nu^3}\cos^3\varphi\nonumber\\&&\hspace{-0.5cm}=0.\label{toruseq}
\end{eqnarray}
If $\nu\rightarrow\infty$ and $P=0$, the above equation degenerates
into Eq.\,(\ref{tubradius}) obeyed by a tubule with radius $\rho$ at
uniform tilting state. If $\nu$ is finite, then Eq.\,(\ref{toruseq})
holds if and only if the coefficients of
$\{1,\cos\varphi,\cos^2\varphi,\cos^3\varphi\}$ vanish. It follows
that $2\lambda \rho^{2}=k_{c}( 4\rho c_{0}-\rho^{2}c_{0}^{2})
-3k_{f}+2h\rho$, $P\rho^{3}=2k_{f}-2k_{c}\rho c_{0}-h\rho$ and
\begin{equation}\nu=\sqrt{\frac{2-k_f/k_c}{1-k_f/k_c}}.\label{torusrad}\end{equation}

The relation between $\nu$, the ratio of the generated radii $r$ and
$\rho$, and $k_f/k_c$ is sketched in Fig.~\ref{torusfig}(b). The
ratio $\nu$ increases with $k_f/k_c$. Especially, $\nu=\sqrt{2}$ for
$k_f/k_c=0$, which leads to the Willmore torus of non-tilting lipid
molecules \cite{oytorus,Seiferttorus}. Since this kind of torus was
observed in the experiments \cite{Bensimon,Rudolphn91,LinHill}, tori
with $\nu>\sqrt{2}$ for $0<k_f/k_c<1$ might also be observed in some
experiments on CLMs.

\section{Chiral lipid membranes with free edges\label{openCLMS}}
In this section, we show the Euler-Lagrange equations and boundary
conditions of CLMs with free edges, and then discuss helical stripes
and twisted ribbons. As we will see, our theoretical result on
twisted ribbons is consistent with the experiment \cite{Oda99}.
\subsection{Euler-Lagrange equations and boundary conditions}
Consider a CLM with an free edge as shown in Fig.~\ref{figframe}(b).
Its free energy can be expressed as
\begin{equation}F=\int G\, dA +\gamma\oint ds,\label{openFE}\end{equation}
where $dA$ is the area element of the membrane and $ds$ the arc
length element of the edge. $\gamma$ represents the line tension of
the edge.

Using the variational method \cite{tzcjpa}, as shown in
Appendix~\ref{appendelopen} we obtain two Euler-Lagrange equations
corresponding to the free energy (\ref{openFE}) as
\begin{equation}
2h(\kappa_{\mathbf{m}}-H)-k_{f}\nabla^{2}\phi=0, \label{EL3}
\end{equation}
and
\begin{eqnarray}
&&2k_{c}\nabla^{2}H+k_{c}( 2H+c_{0}) (
2H^{2}-c_{0}H-2K)-2\lambda H\nonumber\\
&&+h[  \nabla\cdot( \mathbf{m}\nabla\times\mathbf{m}) +\nabla\times(
\mathbf{m}\nabla\cdot\mathbf{m})]\nonumber\\
&&+k_{f}[( \kappa_{\mathbf{v}}-H) \mathbf{v}^{2}-
\nabla\mathbf{v}\colon\nabla\mathbf{e}_3] =0.\label{EL4}
\end{eqnarray}
Additionally, the boundary conditions obeyed by the free edge are
derived as:
\begin{eqnarray}&&\hspace{-0.88cm}k_fv_{b}=0,\label{BC1}\\
&&\hspace{-0.88cm}G+\gamma\kappa_{g}=0,\label{BC2}\\
&&\hspace{-0.88cm}k_{c}(2H+c_{0})-\bar{k}\kappa_{n}-(h/2)\sin2\bar{\phi}=0,\label{BC3}\\
&&\hspace{-0.88cm}\gamma\kappa_{n}-\bar{k}\dot{\tau}_{g}-2k_{c}H_{\mathbf{b}}-h(v_{t}+\dot{\bar{\phi}})\sin2\bar{\phi}
+k_{f}\kappa_{n}v_{t}=0,\label{BC4}\end{eqnarray} where
$\kappa_{n}$, $\tau_{g}$ and $\kappa_{g}$ are the normal curvature,
geodesic torsion, and geodesic curvature of the boundary curve
(i.e., the edge), respectively. $\mathbf{b}$ is in the tangent plane
of the membrane surface and perpendicular to $\mathbf{t}$, the
tangent vector of the boundary curve, as shown in
Fig.~\ref{figframe}(b). $v_b$ and $v_t$ are components of
$\mathbf{v}$ in the direction of $\mathbf{b}$ and $\mathbf{t}$,
respectively. $H_{\mathbf{b}}$ is the direction derivative of $H$
with respect to $\mathbf{b}$. A dot represents the derivative with
respect to $s$. $\bar\phi$ is the angle between $\mathbf{m}$ and
$\mathbf{t}$ at the boundary curve.

It should be noted that Eqs.\,(\ref{EL3}) and (\ref{EL4}) are
equivalent to Eqs.\,(\ref{EL1}) and (\ref{EL2}) with $P=0$.
Eqs.\,(\ref{BC1})--(\ref{BC4}) describe the force and moment balance
relations in the edge. Thus they can also be used for a CLM with
several edges. If $h$ and $k_f$ vanish, the above equations
(\ref{EL3})--(\ref{BC4}) degenerate into the Euler-Lagrange
equations and boundary conditions of open achiral lipid bilayers,
which were fully discussed in
Refs.~\cite{Guven,tzcpre,Umeda05,WangDu06}.

\subsection{Helical stripes}
\begin{figure}[pth!]
\includegraphics[height=6cm]{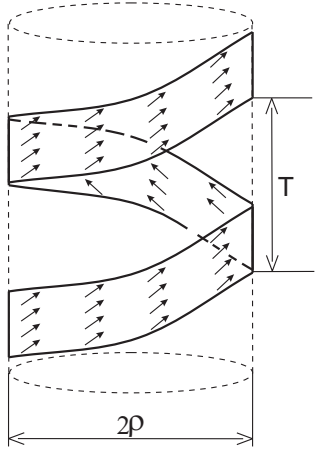}\caption{\label{helixfig}
Helical stripe. Arrows represent the projected directions
$\{\mathbf{m}\}$ of the tilting molecules on the stripe's surface.}
\end{figure}

A helical stripe with pitch $T$ and radius $\rho$ is shown in
Fig.~\ref{helixfig}. In terms of the discussion in
Sec.~\ref{SecHMTS}, we can immediately deduce that a helical stripe
with modulated tilting state is not permitted by the Euler-Lagrange
equations (\ref{EL3}) and (\ref{EL4}) because it can be thought of
as a ribbon wound around a fictitious cylinder. Thus here we only
have to discuss helical stripes with a uniform tilting state, for
which we easily obtain
\begin{eqnarray}&&\phi=\pi/4,\\&&k_c(c_0^2\rho^2-1)+2\lambda\rho^2=0,\label{helixradius}\end{eqnarray}
from the Euler-Lagrange equations. Here $\phi$ is the angle between
$\mathbf{m}$ and the equator of the fictitious cylinder.

In the free edge of the helical stripe, we have $v_b=v_t=0$,
$H=-1/2\rho$, $\bar{\phi}=\pi/4-\psi$, $\tau_\mathbf{m}=-h/2\rho$,
$\kappa_g=0$, $\kappa_n=-\cos^{2}\psi/\rho$,
$\tau_{g}=\sin2\psi/2\rho$, where $\psi\equiv T/2\pi\rho$ is the
pitch angle of the helical stripe. Thus the boundary condition
(\ref{BC1}) is trivial, and Eqs.\,(\ref{BC2})--(\ref{BC4}) can be
transformed into
\begin{eqnarray}
&&(k_c/2)(c_0-1/\rho)^2+\lambda-h/2\rho=0,\label{helixb1}\\
&&k_c(c_0-1/\rho)+\bar{k}\cos^{2}\psi/\rho-(h/2)\cos2\psi=0,\label{helixb2}\\
&&\gamma\cos^{2}\psi/\rho=0.\label{helixb3}
\end{eqnarray}
If $\gamma\neq 0$, there exists only a trivial solution $\psi=\pi/2$
to the above equations (\ref{helixradius})--(\ref{helixb3}). Thus
our theory does not permit genuine helical stripes with free edges
with a uniform tilting state.

\subsection{Twisted ribbons\label{SecTwR}}
\begin{figure}[pth!]
\includegraphics[height=6cm]{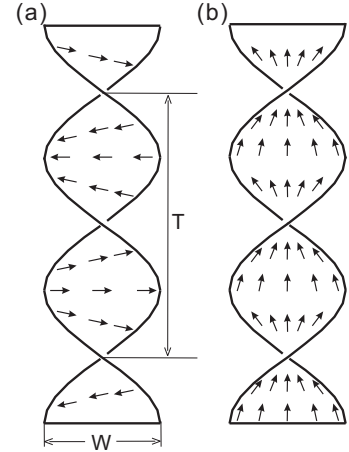}\caption{\label{twistrbfig}
Twisted ribbons: (a) $\mathbf{m}$ is perpendicular to the edges; (b)
$\mathbf{m}$ parallels the edges. Arrows represent the projected
directions $\{\mathbf{m}\}$ of the tilting molecules on the ribbons'
surface.}
\end{figure}

A twisted ribbon with pitch $T$ and width $W$ is shown in
Fig.~\ref{twistrbfig}, which can be expressed as a a vector
$\{u\cos\varphi,u\sin\varphi,\alpha\varphi\}$ with $|u|\leq W/2$,
$|\varphi|<\infty$ and $|\alpha|=T/2\pi$. As shown in
Appendix~\ref{dereqstwist}, Eq.\,(\ref{EL3}) is transformed into
\begin{equation}k_{f}\left(\phi_{uu
}+\frac{u\phi_{u}+\phi_{\varphi\varphi}}{u^{2}+\alpha^{2}}\right)
+\frac{2h\alpha\sin2\phi}{u^{2}+\alpha^{2}}=0.\label{eqtwistr1}
\end{equation} where $\phi$ is the angle between $\mathbf{m}$
and the horizontal.

If we only consider the uniform tilting state, the above equation
requires $\phi=0$ or $\pi/2$. It is easy to see that $\phi=0$
minimizes $-h\int\tau_\mathbf{m}dA$ for $\alpha<0$ while
$\phi=\pi/2$ minimizes $-h\int\tau_\mathbf{m}dA$ for $\alpha>0$ from
Eq.\,(\ref{tautwistr}).Thus we should take $\phi=0$ for $\alpha<0$
and $\phi=\pi/2$ for $\alpha>0$. The former case corresponds to
Fig.~\ref{twistrbfig}(a) where $\mathbf{m}$ is perpendicular to the
edges; the latter corresponds to Fig.~\ref{twistrbfig}(b) where
$\mathbf{m}$ is parallel to the edges. As shown in
Appendix~\ref{dereqstwist}, both for $\phi=0$ and $\pi/2$,
Eq.\,(\ref{EL4}) is transformed into
\begin{equation}k_cc_0\alpha^2/(u^2+\alpha^2)^2=0,\label{eqtwistr2}\end{equation}
which requires $c_0=0$ for non-vanishing $\alpha$. Among the
boundary conditions (\ref{BC1})--(\ref{BC4}), only Eq.\,(\ref{BC2})
is nontrivial, which gives
\begin{equation}\lambda(1+x^{2})  \alpha^{2}-(h-\gamma x)|\alpha|+\frac{2\bar
{k}+k_{f}x^{2}}{2(1+x^{2})}=0,\label{BC22}\end{equation} with
$x\equiv W/2|\alpha|$.

\begin{figure}[pth!]
\includegraphics[width=7cm]{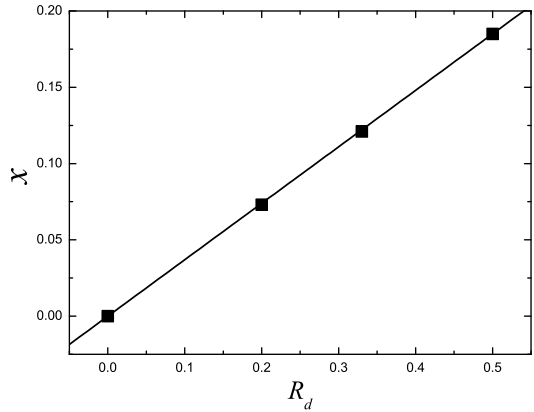}\caption{\label{fignature}
Relation between $x=W/2|\alpha|$ and $R_d$. The solid line is the
fitting curve $x=0.37R_d$ and the dots are the experimental data in
Ref.~\cite{Oda99}.}
\end{figure}

Guided by the experimental data \cite{Oda99}, we may assume $h$ to
be proportional to $R_d$, the relative concentration difference of
the left- and right-handed enantiomers in the experiment, i.e.,
$h=h_0 R_d$ with a constant $h_0$. In terms of the experimental
data, $|\alpha|\rightarrow\infty$ for $R_d\rightarrow 0$. Thus
Eq.\,(\ref{BC22}) requires $\lambda=0$ and then
\begin{equation}|\alpha|=(2\bar{k}+k_{f}x^{2})/2(h_0R_d-\gamma x)(1+x^{2}).\label{alfeqn}\end{equation}
To determine the relation between $x$ and $R_d$, we minimize the
average energy per area with respect to $|\alpha|$ for given $W$.
The average energy per area is calculated as
\begin{widetext}
\begin{equation}\bar{F}=\frac{(2\bar{k}-k_{f})x+(k_{f}-2h_0R_d|\alpha|)\sqrt{1+x^{2}}\mathrm{arcsinh}%
x+2\gamma|\alpha|(1+x^{2})}{\alpha^{2}\sqrt
{1+x^{2}}(\mathrm{arcsinh}x+x\sqrt
{1+x^{2}})}.\label{fretwistr}\end{equation}\end{widetext}Minimizing
it with respect to $|\alpha|$ and using Eq.\,(\ref{alfeqn}), we
obtain
\begin{equation}x=\beta R_d+O(R_d^3),\label{ratioh}\end{equation}
with $\beta\equiv 3h_0/4\gamma$. This relation reveals that the
ratio between the width and pitch of the ribbons is proportional to
the relative concentration difference of left- and right-handed
enantiomers in the low relative concentration difference region.
Eq.\,(\ref{ratioh}) fits well with the experimental data with
parameter $\beta=0.37$ as shown in Fig.~\ref{fignature}.

\section{Conclusion and discussion\label{summary}}
In this paper, we have focused on a concise free energy from which
the Euler-Lagrange equations of CLMs without free edges and
Euler-Lagrange equations as well as boundary conditions of CLMs with
free edges can be derived analytically. Their implications for
extant and future experiments can be summarized as follows.

(i) Our theory predicts the same probability for left- and
right-handed spherical vesicles existing in solution. Thus no
evident circular dichroism signal should be observed, which is
consistent with the experiment \cite{Schnur94}.

(ii) The radius of a tubule at uniform tilting state satisfies
Eq.\,(\ref{tubradius}). It reveals that the radius is independent of
molecular chirality, which is consistent with the experiment
\cite{Spector98} where the tubular radius is insensitive to the
proportion of left- and right-handed DC$_{8,9}PC$ in the tubule. We
find that the projected direction of the molecules on the tubular
surface departs 45$^\circ$ from the equator of the tubule with the
uniform tilting state, which is in good agreement with the
experiment \cite{Zhaoy05}.

(iii) Tubules with helically modulated tilting state are not
admitted by the Euler-Lagrange equation (\ref{EL2}) for
non-vanishing $h$. The orientational variation of the tilting lipid
molecules breaks the force balance along the normal direction of the
tubule. Thus a tubule with helically modulated tilting state is not
an equilibrium structure within our theoretical framework.

(iv) Helical ripples in tubules are equilibrium structures. The
pitch angle of helical ripples is estimated as about 0$^\circ$ and
35$^\circ$ for $k_f\simeq k_c$, which are close to the experimental
values 5$^\circ$ and 28$^\circ$ observed by Fang's group
\cite{Zhaoy06}.

(v) Tori with a ratio of generated radii larger than $\sqrt{2}$ are
predicted by our theory, Eq.\,(\ref{torusrad}), which have not yet
been observed in the experiments.

(vi) Helical stripes with free edges at either uniform tilting state
or helical modulated tilting state are not possible equilibrium
configurations.

(vii) Twisted ribbons satisfy Euler-Lagrange equations (\ref{EL3}),
(\ref{EL4}) and boundary conditions (\ref{BC1})--(\ref{BC4}). The
ratio between the width and pitch of the ribbons is proportional to
the relative concentration difference of left- and right-handed
enantiomers in the low relative concentration difference region,
which is in good agreement with the experiment \cite{Oda99}.

Finally, we have to list a few open problems which should be
addressed in the future work.

(i) Within the present theory, we cannot give a simple explanation
for the experiments on CLMs of pure cholesterol
\cite{Chung93,Zastavker} where helical stripes with pitch angles
$11^\circ$ and $54^\circ$ are usually observed. It has recently been
found that the cholesterol helical stripes have good crystal
structure \cite{Khaykovich07}, which is out of the range of our
theory. There might be two possible ramifications based on the
present theory: One would be to include anisotropic bending effects
\cite{ChenCM99} in the free energy density (\ref{energy2}); Another
one would be to consider a line tension $\gamma$ depending on the
angle between the directions of the tilting and the free edges
\cite{Sarasij02} in the free energy (\ref{openFE}).

(ii) We cannot yet interpret the twisted ribbon-to-tubule transition
with increasing the relative concentration difference of the left-
and right-handed enantiomers reported in the recent experiment by
Oda's group \cite{OdaJACS07}. A possible reason is that the
parameters except $h$ in our theory are independent of the relative
concentration difference \cite{odacom}. Additionally, the ribbons
and tubules observed in this experiment are usually multi-bilayer
structures \cite{odacom}. Then a decoupling effect
\cite{Seifert96,Fournier} between neighbor bilayers might occur. An
extended theory including these two factors might be required to
investigate the mechanism responsible for the transition.

(iii) We have ignored the effect of singular points in CLMs.
Although the features of singular points in 2D planar films or
achiral nematic spherical, torus vesicles and other manifolds have
been fully investigated in previous literature
\cite{Sarasij02,Langer92,PetteyPRE99,VitelliPRE06,Powers07}, it
remains a challenge to study the properties of singular points in
CLMs \cite{rmkspoint}.

\section*{Acknowledgement}
ZCT is grateful to Z. C. Ou-Yang, J. Fang and R. Oda for instructive
discussions and to the Alexander von Humboldt foundation for
financial support.

\appendix
\section{\label{Appdmfm}Brief introduction to moving frame method and exterior differential forms}
If we take a frame $\{\mathbf{e}_1,\mathbf{e}_2,\mathbf{e}_3\}$ at
any point $\mathbf{r}$ on a surface as shown in Fig.~\ref{figframe},
then the infinitesimal tangential vector at $\mathbf{r}$ is
expressed as
\begin{equation}d\mathbf{r}=\omega_1\mathbf{e}_1+\omega_2\mathbf{e}_2\label{drframe}\end{equation} and the
difference of frame between at points $\mathbf{r}+d\mathbf{r}$ and
$\mathbf{r}$ is denoted as
\begin{equation}d\mathbf{e}_i=\omega_{ij}\mathbf{e}_j,\quad (i=1,2,3),\label{diffei}\end{equation}
where $\omega_1$, $\omega_2$, and $\omega_{ij}=-\omega_{ji},
(i,j=1,2,3)$ are 1-forms \cite{tzcjpa,Chernbook}. The repeated
subscripts in Eq.\,(\ref{diffei}) and the following contents
represent the Einstein summation convention. With these 1-forms, the
structure equations of a surface can be expressed as
\cite{tzcjpa,Chernbook}
\begin{equation}
\left\{\begin{array}{l}d\omega_1=\omega_{12}\wedge\omega_2,\\
d\omega_2=\omega_{21}\wedge\omega_1,\\
\label{domgaij} d\omega_{ij}=\omega_{ik}\wedge\omega_{kj}\quad
(i,j=1,2,3),\end{array}\right. \label{structur}\end{equation} and
\begin{equation}
\left(\begin{array}{l}\omega_{13}\\
\omega_{23}\end{array}\right)=\left(\begin{array}{cc}a&b\\
b&c\end{array}\right)\left(\begin{array}{l}\omega_{1}\\
\omega_{2}\end{array}\right),\label{omega13}\end{equation} where the
symbol `$\wedge$' expresses the wedge product between differential
forms and `$d$' is the exterior differential operator
\cite{tzcjpa,Chernbook}. The
matrix $\left(\begin{array}{cc}a&b\\
b&c\end{array}\right)$ is called the curvature matrix which is
related to the mean and gaussian curvature by
\begin{equation}2H=a+c,\quad K=ac-b^2.\label{HKgeneral}\end{equation}

For a unit vector
$\mathbf{m}=\cos\phi\mathbf{e}_1+\sin\phi\mathbf{e}_2$, the normal
curvature along the direction of $\mathbf{m}$ can be expressed as
\cite{tzcjpa}
\begin{equation}
\kappa_{\mathbf{m}}=a\cos^2\phi+2b\cos\phi\sin\phi+c\sin^2\phi.\label{normalcur}\end{equation}
The normal curvature along the direction of an arbitrary vector
$\mathbf{v}$ can be expressed as
\begin{equation}
\kappa_{\mathbf{v}}=(av_1^2+2bv_1v_2+cv_2^2)/\mathbf{v}^2,\label{norcurinv}\end{equation}
where $v_1$ and $v_2$ are the components of $\mathbf{v}$ in the
directions of $\mathbf{e}_1$ and $\mathbf{e}_2$.

In our following derivations, several relations between vector forms
and differential forms are used frequently. For convenience, we list
them below.
\begin{eqnarray}
&&(\nabla\times \mathbf{m})dA=d(\mathbf{m}\cdot d\mathbf{r}),\label{curlm}\\
&&(\nabla\cdot \mathbf{m})dA=d(\ast\mathbf{m}\cdot d\mathbf{r}),\label{divgm}\\
&&\tau_{\mathbf{m}}dA=\mathbf{m}\cdot
d\mathbf{e}_3\wedge\mathbf{m}\cdot d\mathbf{r},\label{daudfine}\\
&&\kappa_{\mathbf{m}}dA=-\mathbf{m}\cdot
d\mathbf{e}_3\wedge\ast\mathbf{m}\cdot d\mathbf{r},\\
&&\nabla \phi \cdot d\mathbf{r}=d\phi,\label{nabdphi}\\
&&(\nabla^2\phi) dA=d\ast d\phi,\label{laplphi}\\
&&\mathbf{S}\cdot d\mathbf{r}=-\omega_{12}\label{Sandomega12},\\
&&(\nabla\cdot\mathbf{S})dA=-d\ast\omega_{12},\label{nabdS}\\
&&(\nabla\times\mathbf{S})dA=-d\omega_{12}=KdA,\\
&&\mathbf{v}\cdot d\mathbf{r}=d\phi+\omega_{12},\label{vdefin1}\\
&&\mathbf{v}^2dA=\mathbf{v}\cdot
d\mathbf{r}\wedge\ast\mathbf{v}\cdot
d\mathbf{r},\label{vdefin2}\\
&&(\nabla\mathbf{v}\colon\nabla\mathbf{e}_3)dA=d(\mathbf{v}\cdot
d\mathbf{e}_3),\label{nabvnabe3}
\end{eqnarray}
where $dA=\omega_1\wedge\omega_2$ is the area element. $\mathbf{m}$
is a unit vector and $\phi$ is the angle between $\mathbf{m}$ and
$\mathbf{e}_1$. $\mathbf{S}$ is the spin connection and
$\mathbf{v}\equiv\nabla\phi-\mathbf{S}$. $\ast$ is the Hodge star
operator \cite{tzcjpa,Westenholzbk} which satisfies $\ast
\omega_1=\omega_2$ and $\ast\omega_2=-\omega_1$.

Using Eqs.\,(\ref{HKgeneral}), (\ref{norcurinv}), (\ref{curlm}),
(\ref{divgm}) and (\ref{vdefin1}), we can prove
\begin{equation}(\kappa_{\mathbf{v}}-H)  \mathbf{v}^{2}=(v_{1}^{2}-v_{2}^{2})(a-c)/2  +2bv_{1}v_{2},\label{kvmHgen}\end{equation}
and
\begin{eqnarray}&&[\nabla\cdot(
\mathbf{m}\nabla\times\mathbf{m}) +\nabla\times(
\mathbf{m}\nabla\cdot\mathbf{m}) ]  dA\label{ntndmp} \\
&&\hspace{-0.4cm}=d[(v_{2}\cos2\phi- v_{1}\sin2\phi)  \omega _{1}+(
v_{1}\cos2\phi+v_{2}\sin2\phi ) \omega_{2}],\nonumber
\end{eqnarray}
where $v_1$ and $v_2$ are the components of $\mathbf{v}$ in the
directions of $\mathbf{e}_1$ and $\mathbf{e}_2$.

We suggest the reader refer first to Ref.~\cite{tzcjpa} and some
textbook on the calculus with the moving frame method before going
on if he is interested in the mathematical details. In writing the
following contents we assume that the reader has been familiar with
the skill in Ref.~\cite{tzcjpa} and good at calculus with the moving
frame method.

\section{Derivation of the Euler-Lagrange equations of CLMS without free edges\label{appendelclose}}
Here the CLMS without free edges will be derived through the
variational method developed in Ref.~\cite{tzcjpa} with the aid of
the moving frame method and exterior differential forms, which can
highly simplify the calculus of variation.

\subsection{Variation respect to $\phi$}
Assume $\delta\phi=\Xi$. Through simple calculations, we arrive at
\begin{equation}\delta\tau_{\mathbf{m}}=2(H-\kappa_{\mathbf{m}})\Xi,\end{equation}
and
\begin{equation}\delta(\mathbf{v}^{2}dA)=2d\Xi\wedge\ast(d\phi+\omega_{12}).\end{equation}
Combining with the above two equations and using the integral by
parts and Stokes' theorem, we derive
\begin{equation}
\delta F=\int[ 2h(\kappa_{\mathbf{m}}-H)-k_{f}\nabla^{2}\phi] \Xi
dA+k_{f} \oint \Xi\ast\mathbf{v}\cdot d\mathbf{r},\label{varyphi}
\end{equation}
where the second term is the integral along the boundary curve,
which vanishes for CLMS without free edges. Since $\Xi$ is an
arbitrary function, from $\delta F=0$ we derive Euler-Lagrange
equation (\ref{EL1}). We have used the assumption $\nabla\cdot
\mathbf{S}=0$ when we write Eq.\,(\ref{varyphi}). This assumption is
indeed satisfied in our discussions on all configurations except
ripples in tubules. Thus $\nabla^{2}\phi$ should be replaced with
$\nabla^{2}\phi-\nabla\cdot \mathbf{S}$ in Eq.\,(\ref{varyphi}) as
well as Eq.\,(\ref{EL1}) when this condition is not met.

\subsection{Variation respect to the deformation of the surface}
In this subsection, let us denote $F_0$ as the functional
(\ref{closedFE}) with $h$ and $k_f$ vanishing, and define the
additional functional
\begin{equation}F_{ad}=\int \left(\frac{k_{f}}{2}\mathbf{v}
^{2}-h\tau_{\mathbf{m}}\right)dA.\label{Fadd}\end{equation}

Any small deformation of a CLM without free edges can always be
achieved from small normal displacement $\Omega_{3}$ at each point
$\mathbf{r}$ in the surface. That is,
$\delta\mathbf{r}=\Omega_{3}\mathbf{e}_{3}$. The frame is also
changed because of the deformation of the surface, which is denoted
as
\begin{equation}\delta\mathbf{e}_i=\Omega_{ij}\mathbf{e}_j,\quad (i=1,2,3),\label{varyei}\end{equation}
where $\Omega_{ij}=-\Omega_{ji},(i,j=1,2,3)$ corresponds to the
rotation of the frame due to the deformation of the surface.

$\delta F_0$ was first dealt with in Ref.~\cite{oy90} which gives
\begin{eqnarray}\delta F_0&=&\int (2k_{c}\nabla^{2}H-2\lambda
H+P)\Omega_{3}dA\nonumber\\&+&\int k_{c}( 2H+c_{0}) (
2H^{2}-c_{0}H-2K)\Omega_{3}dA.\end{eqnarray}

Following Ref.~\cite{tzcjpa}, and considering
Eqs.\,(\ref{daudfine}), (\ref{vdefin1}) and (\ref{vdefin2}), through
somewhat involved calculations, we can derive
\begin{eqnarray}&& \delta( \tau_{\mathbf{m}}dA)
=2\Omega_{12}(H-\kappa _{\mathbf{m}}) dA + \nabla
\cdot\mathbf{m}d\Omega_{3}\wedge\mathbf{m}\cdot
d\mathbf{r}\nonumber\\&&\hspace{1.4cm}+\mathbf{m}\cdot
d\mathbf{r}\wedge d( d\Omega_{3}\wedge\ast\mathbf{m}\cdot
d\mathbf{r}/dA) ,\label{vartaum}\\
&&\delta(\mathbf{v}^{2}dA) =2d\Omega_{12}\wedge\ast(
d\phi+\omega_{12})\nonumber\\
&&\hspace{1.4cm}+2d\Omega_{3}\wedge\mathbf{v}\cdot
{d}\mathbf{e}_3+2( \kappa_{\mathbf{v}}-H) \mathbf{v}^{2}\Omega
_{3}dA.\label{varkappam}\end{eqnarray} Combining with the above two
equations and using the integral by parts and Stokes' theorem, we
derive
\begin{eqnarray}
&&\delta F_{ad}=\int h[  \nabla\cdot(
\mathbf{m}\nabla\times\mathbf{m}) +\nabla\times(
\mathbf{m}\nabla\cdot\mathbf{m})]\Omega_{3}dA\nonumber\\
&&\hspace{0.85cm}+\int k_{f}[( \kappa_{\mathbf{v}}-H)
\mathbf{v}^{2}- \nabla\mathbf{v}\colon\nabla\mathbf{e}_3]
\Omega_{3}dA,\label{varyOmega3}
\end{eqnarray}
Then the Euler-Lagrange equation (\ref{EL2}) follows from $\delta
F=\delta F_0+\delta F_{ad}=0$ because $\Omega_{3}$ is an arbitrary
function.

\section{Derivation of the Euler-Lagrange equations and boundary conditions of CLMS with free edges\label{appendelopen}}
A CLM with an free edge can be described as a surface with an
boundary curve shown in Fig.~\ref{figframe}B.

\subsection{Variation respect to $\phi$}
Let $\delta\phi=\Xi$, we still have Eq.\,(\ref{varyphi}) whose
second term is transformed into
\begin{equation}-k_f\oint v_b\Xi ds\end{equation}
with $v_b=\mathbf{v}\cdot \mathbf{b}$. Thus from $\delta F=0$ we can
derive the Euler-Lagrange equation (\ref{EL3}) and boundary
condition (\ref{BC1}).

\subsection{Variation respect to the deformation of the surface}
Because one can select an arbitrary frame
$\{\mathbf{r};\mathbf{e}_1,\mathbf{e}_2,\mathbf{e}_3\}$, here we
take it such that $\mathbf{e}_1$ and $\mathbf{e}_2$ align with
$\mathbf{t}$ and $\mathbf{b}$ in the boundary curve. Then any small
deformation of a CLM with an free edge can always be expressed as
the linear superposition of small normal displacement $\Omega_{3}$
and tangent displacement $\Omega_{2}$ along $\mathbf{e}_2$ at each
point $\mathbf{r}$ in the surface.

First, we consider the in-plane deformation mode
$\delta\mathbf{r}=\Omega_{2}\mathbf{e}_{2}$. The change of the frame
is still denoted as Eq.\,(\ref{varyei}). In terms of
Ref.~\cite{tzcpre}, we have $\delta \oint ds=- \oint \kappa_g
\Omega_{2}ds$ where $\kappa_g$ is the geodesic curvature of the
boundary curve. Additionally, we can derive
\begin{equation}\delta\int G dA=-\oint G \Omega_{2}ds.\label{varyOmega2G}\end{equation}
Although the derivation of the above equation is somewhat involved,
its physical meaning is quite clear. Under the displacement
$\Omega_{2}$ along $\mathbf{e}_2$, $G$ is similar to the surface
tension and the area element near the boundary decreases by
$\Omega_{2}ds$. Thus the variation of $\int G dA$ gives
Eq.\,(\ref{varyOmega2G}) and then we obtain the boundary condition
(\ref{BC2}) from $\delta F=0$.

Second, we consider the out-of-plane deformation mode
$\delta\mathbf{r}=\Omega_{3}\mathbf{e}_{3}$. Let us denote $F_0$ as
the functional (\ref{openFE}) with $h$ and $k_f$ vanishing, and
define the additional functional as Eq.\,(\ref{Fadd}). $\delta F_0$
is fully discussed in Ref.~\cite{tzcpre} as
\begin{eqnarray}&&\hspace{-0.4cm}\delta F_0=\int [k_{c}(
2H+c_{0}) ( 2H^{2}-c_{0}H-2K)-2\lambda H]\Omega_{3}dA\nonumber\\
&&\hspace{-0.4cm}+\int 2k_{c}(\nabla^{2}H)\Omega_{3} dA
-\oint[k_{c}(2H+c_{0})-\bar{k}\kappa_{n}]\Omega_{23}ds\nonumber\\
&&\hspace{-0.4cm}-\oint[-2k_{c}\partial
H/\partial\mathbf{b}+\gamma\kappa_{n}-\bar {k}\dot{\tau}_{g}]\Omega
_{3}ds.\label{BCF0}\end{eqnarray} Here the only difference is that
the sign of $\bar{k}$ adopted in the present paper is opposite to
that in Ref.~\cite{tzcpre}. Similar to the derivation of
Eq.\,(\ref{varyOmega3}) from Eqs.\,(\ref{vartaum}) and
(\ref{varkappam}) by using the integral by parts and Stokes'
theorem, we have \begin{eqnarray}\delta F_{ad}&=&\int h[\nabla\cdot(
\mathbf{m}\nabla\times\mathbf{m}) +\nabla\times(
\mathbf{m}\nabla\cdot\mathbf{m})]\Omega_{3}dA\nonumber\\&+&\int
k_{f}[( \kappa_{\mathbf{v}}-H) \mathbf{v}^{2}-
\nabla\mathbf{v}\colon\nabla\mathbf{e}_3] \Omega_{3}dA\nonumber\\
&+&\oint[h(v_{t}+\dot{\bar{\phi}})\sin2\bar{\phi}-k_{f}\kappa_{n}v_{t}]\Omega_{3}ds
\nonumber\\&+&\oint(h/2)\sin2\bar{\phi}\Omega_{23}ds,\label{BCFadd}\end{eqnarray}
where $v_t=\mathbf{v}\cdot \mathbf{t}$. In the above equations
(\ref{BCF0}) and (\ref{BCFadd}), $\Omega_{3}$ represents the
arbitrary small displacement of point on the surface along
$\mathbf{e}_3$ and $\Omega_{23}$ is the arbitrary small rotation of
$\mathbf{e}_3$ around $\mathbf{t}$ at the edge. Thus $\delta F=0$
will give the Euler-Lagrange equation (\ref{EL4}) and boundary
conditions (\ref{BC3}) and (\ref{BC4}).

\section{\label{deriveqshelic}Derivation of equations (\ref{stripe01}) and (\ref{stripe02}) for tubules with nonuniform tilting state in Section \ref{SecHMTS}}
For a cylindrical surface, take a frame such that $\mathbf{e}_1$,
$\mathbf{e}_2$ and $\mathbf{e}_3$ are along the circumferential,
axial and radial directions, respectively. Then we have
\begin{eqnarray}
&&\omega_{1}=ds,\omega_{2}=dz\\
&&a=-1/\rho,b=c=0.\label{abccylind}\end{eqnarray} Thus
\begin{equation}2H=a+c=-1/\rho,K=ac-b^2=0.\label{HKcylind}\end{equation}
Because $dds=0$ and $ddz=0$, using the structure equation
(\ref{structur}), we have $\omega_{12}=0$. Considering
Eq.\,(\ref{Sandomega12}), we obtain the spin connection
\begin{equation}\mathbf{S}=0,\end{equation}
and then
\begin{equation}\mathbf{v}=\nabla\phi=\phi_{s}\mathbf{e}%
_{1}+\phi_{z}\mathbf{e}_{2}.\label{vfieldcylind}\end{equation} From
Eqs.\,(\ref{normalcur}) and (\ref{laplphi}), we have
\begin{eqnarray}&&\kappa_{\mathbf{m}}=-\cos^{2}\phi/\rho,\\
&&\nabla^2\phi=\phi_{ss}+\phi_{zz}.\end{eqnarray} Substituting the
above two equations and Eq.\,(\ref{HKcylind}) into the
Euler-Lagrange equation (\ref{EL1}), we can derive
Eq.\,(\ref{stripe01}) in Sec.~\ref{SecHMTS}.

Additionally, using Eqs.\,(\ref{diffei}), (\ref{omega13}) and
(\ref{vfieldcylind}), we derive
\begin{equation}\mathbf{v}\cdot d\mathbf{e}_3=(\phi_s/\rho)ds.\end{equation}
From Eq.\,(\ref{nabvnabe3}) we can derive
\begin{equation}\nabla\mathbf{v}\colon\nabla\mathbf{e}_3=-\phi_{sz}/\rho.\label{ve3cylind}\label{nvne3cylind}\end{equation}
In this derivation, one should note that $dA=ds\wedge dz=-dz\wedge
ds$. From Eqs.\,(\ref{kvmHgen}), (\ref{ntndmp}), (\ref{abccylind})
and (\ref{vfieldcylind}), we can obtain
\begin{equation}(\kappa_{\mathbf{v}}-H)  \mathbf{v}^{2}=(\phi_z^{2}-\phi_s^{2})/2\rho,\end{equation}
and
\begin{eqnarray}&&\nabla\cdot(
\mathbf{m}\nabla\times\mathbf{m}) +\nabla\times(
\mathbf{m}\nabla\cdot\mathbf{m})\label{ntndmp2}\\
&&\hspace{-0.4cm}=2\sin2\phi( \phi_{z}^{2}-\phi_{s}^{2}+\phi_{sz})
+\cos2\phi(
\phi_{ss}-\phi_{zz}+4\phi_{z}\phi_{s}).\nonumber\end{eqnarray}

Substituting Eqs.\,(\ref{HKcylind}) and
(\ref{ve3cylind})--(\ref{ntndmp2}) into the Euler-Lagrange equation
(\ref{EL2}), we can obtain Eq.\,(\ref{stripe02}) in
Sec.~\ref{SecHMTS}.

\section{\label{dereqsripp}Derivation of equations (\ref{eqripp1}) and (\ref{eqripp2}) for helical ripples in Section \ref{Secripple}}
For the surface expressed as vector form
\begin{equation}
\mathbf{r}=\{\rho(1+y) \cos(s/\rho) ,\rho(1+y)  \sin(s/\rho)
,z\}\end{equation} with $|y|\ll 1$, the first and second fundamental
forms of the surface can be calculated as
\begin{eqnarray}&&\hspace{-0.9cm}I=(1+2y)ds^2+dz^2,\\
&&\hspace{-0.9cm}II=-\frac{1}{\rho}(  1+y-\rho^{2}y_{ss})
ds^{2}+2\rho y_{sz}dsdz+\rho y_{zz}dz^{2}\end{eqnarray} up to the
order of $O(y)$, respectively. The same order is kept in the
following expressions in this section. In terms of the
correspondence relations $I=\omega_1^2+\omega_2^2$ and
$II=a\omega_1^2+2b\omega_1\omega_2+\omega_2^2$, we have
\begin{eqnarray}&&\omega_1=(1+y)ds,\omega_2=dz,\\
&&a=-( 1-y-\rho^{2}y_{ss})/{\rho}, b=\rho y_{sz},c=\rho y_{zz},\\
&&2H=-(1-y-\rho^{2}y_{ss}-\rho^{2}y_{zz})/{\rho},
K=-y_{zz}.\label{HKripples}
\end{eqnarray}
$2H$ is a function of $s$ and $z$, using $\nabla^2(2H)dA=d\ast
d(2H)$, we can derive
\begin{equation}\nabla^{2}(2H)   =(y_{ss}
+y_{zz}+\rho^{2}y_{ssss}+2\rho^{2}y_{zzss}+\rho^{2}y_{zzzz})/\rho\end{equation}

Because $d\omega_1=y_zdz\wedge ds$ and $d\omega_2=ddz=0$, using the
structure equation (\ref{structur}), we have
$\omega_{12}=-y_z\omega_1$. Considering Eqs.\,(\ref{Sandomega12})
and (\ref{nabdS}), we have
\begin{equation}\mathbf{S}=y_z\mathbf{e}_1,\nabla\cdot\mathbf{S}=y_{zs}.\end{equation}

Using Eqs.\,(\ref{nabdphi}) and (\ref{laplphi}), we obtain
\begin{eqnarray}&&\nabla\phi=\phi_{s}(1-y)\mathbf{e}_{1}+\phi_{z}\mathbf{e}_{2}\\
&&\nabla^{2}\phi
=\phi_{ss}+\phi_{zz}+\phi_{z}y_{z}-\phi_{s}y_{s}-2y\phi_{ss}.\end{eqnarray}
and then
\begin{equation}\mathbf{v}=\nabla\phi-\mathbf{S}=[\phi_{s}(1-y)-y_z]\mathbf{e}_{1}+\phi_{z}\mathbf{e}_{2}.\label{vfieldripp}\end{equation}

From Eqs.\,(\ref{diffei}), (\ref{omega13}), (\ref{nabvnabe3}),
(\ref{kvmHgen}), and (\ref{vfieldripp}), we can derive
\begin{eqnarray}&&\hspace{-0.5cm}\nabla\mathbf{v}\colon\nabla\mathbf{e}_{3}=-\phi_{sz}/\rho+(y_{z}\phi_{s}+2y\phi_{sz}+y_{zz})/\rho,\\
&&\hspace{-0.5cm}(  \kappa_{\mathbf{v}}-H)
\mathbf{v}^{2}=(\phi_{z}^{2}-\phi_{s}^{2})/{2\rho}+\phi_{s}(
y\phi_{s}+y_{z})/{\rho}\nonumber\\
&&\hspace{-0.7cm}+2\rho
y_{sz}\phi_{s}\phi_{z}+(y+\rho^{2}y_{ss}-\rho^{2}y_{zz})(\phi_{s}^{2}-\phi_{z}^{2})/2\rho.
\end{eqnarray}
Since we only consider the case of $\bar{h}\ll 1$, up to the order
$O(\bar{h})$, we obtain
\begin{eqnarray}
&&\hspace{-0.4cm}\bar{h}(\kappa_\mathbf{m}-H)=-(\bar{h}/2\rho)\cos2\phi,\\
&&\hspace{-0.4cm}\bar{h}[\nabla\cdot(
\mathbf{m}\nabla\times\mathbf{m}) +\nabla\times(
\mathbf{m}\nabla\cdot\mathbf{m})]\label{ntndmrip}\\
&&\hspace{-0.4cm}=\bar{h}[2(\phi_{z}^{2}-\phi_{s}^{2}+\phi_{sz})\sin2\phi
+( \phi_{ss}-\phi_{zz}+4\phi_{z}\phi_{s})\cos2\phi]\nonumber
\end{eqnarray}
by using Eqs.\,(\ref{normalcur}) and (\ref{ntndmp2}).

Substituting the above equations (\ref{HKripples})--(\ref{ntndmrip})
into the Euler-Lagrange equations (\ref{EL1}) and (\ref{EL2}), we
can derive the basic equations (\ref{eqripp1}) and (\ref{eqripp2})
under the assumption $c_0=0$, $\lambda=0$, $k_f\simeq k_c$ in
Sec.~\ref{Secripple}.

\section{\label{dereqstorus}Derivation of equations (\ref{phitorus}) and (\ref{toruseq}) for tori in Section \ref{Sectorus}}
A torus can be expressed as a vector form
\begin{equation}
\mathbf{r}=\{(r+\rho\cos\varphi)\cos\theta,(r+\rho\cos\varphi)\sin\theta,\rho\sin\varphi\}\label{vectorus}
\end{equation}
with $r>\rho$. A frame $\{\mathbf{e}_1,\mathbf{e}_2,\mathbf{e}_3\}$
is taken as
\begin{equation}\left\{\begin{array}{l}
\mathbf{e}_{1}=\{-\sin \theta ,\cos \theta ,0\},\\
\mathbf{e}_{2}=\{-\sin \varphi \cos \theta ,-\sin \varphi \sin
\theta ,\cos \varphi \}, \\
\mathbf{e}_{3}=\{\cos \theta \cos \varphi ,\sin \theta \cos \varphi
,\sin \varphi \}.
\end{array}\right.\label{framtorus}\end{equation}

By using Eqs.\,(\ref{drframe}), (\ref{vectorus}) and
(\ref{framtorus}), we can derive
\begin{equation}\omega_1=(r+\rho\cos \varphi )d\theta,\omega_2=\rho d\varphi.\end{equation}

From Eq.\,(\ref{diffei}) and (\ref{framtorus}), we obtain
\begin{eqnarray}
&&\omega_{12}=[{\sin \varphi }/{(r+\rho\cos \varphi )}]\omega_1,\label{omega12torus}\\
&&\omega_{13}=-[{\cos \varphi }/{(r+\rho\cos \varphi )}]\omega_1,\label{omega13torus}\\
&&\omega_{23}=-(1/\rho)\omega_2.\label{omega23torus}
\end{eqnarray}
Considering Eqs.\,(\ref{Sandomega12}), (\ref{nabdS}) and
(\ref{omega12torus}), we have
\begin{eqnarray}&&\mathbf{S}=-[{\sin \varphi }/{(r+\rho\cos \varphi )}]\mathbf{e}_1,\\
&&\nabla\cdot\mathbf{S}=0.
\end{eqnarray}
Additionally, (\ref{omega13}), (\ref{omega13torus}) and
(\ref{omega23torus}) give
\begin{equation}a=-{\cos \varphi }/{(r+\rho\cos \varphi )},b=0,c=-1/\rho.\label{abctorus}\end{equation}
Thus by considering Eqs.\,(\ref{torsionexp}), (\ref{HKgeneral}),
(\ref{normalcur}), and (\ref{laplphi}), we obtain
\begin{eqnarray}&&\hspace{-0.7cm}2H=-\frac{r+2\rho\cos\varphi}{\rho( r+\rho\cos\varphi)},K=\frac{\cos\varphi}{\rho(r+\rho\cos\varphi)},\label{HKtorus}\\
&&\hspace{-0.7cm}\kappa_{\mathbf{m}}=-\frac{\cos^{2}\phi\cos\varphi}{r+\rho\cos\varphi}-\frac{\sin^{2}\phi}{\rho},\\
&&\hspace{-0.7cm}\tau_{\mathbf{m}}=-\frac{r\sin2\phi}{2\rho(r+\rho\cos\varphi)},\label{tautours}\\
&&\hspace{-0.7cm}\nabla^{2}\phi
=\frac{\phi_{\theta\theta}}{(r+\rho\cos\varphi)^{2}}+\frac{\phi_{\varphi\varphi}}{\rho^{2}}-\frac{\sin\varphi\phi_{\varphi}}{\rho(
r+\rho\cos\varphi)}.\end{eqnarray} Substituting the above four
equations into Eq.\,(\ref{EL1}), we can derive Eq.\,(\ref{phitorus})
in Sec.~\ref{Sectorus}.

If $\phi=-\pi/4$, then $\nabla\phi=0$, and
\begin{eqnarray}&&\mathbf{v}\equiv\nabla\phi-\mathbf{S} =[{\sin \varphi }/{(r+\rho\cos \varphi )}]\mathbf{e}_1,\label{vfieldtorus}\\
&& \mathbf{v}\cdot
d\mathbf{e}_3=v_1\omega_{31}=[\sin2\varphi/2(r+\rho\cos\varphi)]d\theta,
\end{eqnarray}
where $v_1$ is the component of $\mathbf{v}$ in the direction of
$\mathbf{e}_1$. Thus from Eq.\,(\ref{nabvnabe3}) we can derive
\begin{equation}\nabla\mathbf{v}\colon\nabla\mathbf{e}_{3}=-(r\cos2\varphi
+\rho\cos^{3}\varphi)/\rho(r+\rho\cos\varphi)^{3}.\label{nabvnabe3trs}
\end{equation}
Using Eqs.\,(\ref{kvmHgen}), (\ref{ntndmp}), (\ref{abctorus}) and
(\ref{vfieldtorus}), we can obtain
\begin{equation}\hspace{-1.2cm} (\kappa_{\mathbf{v}}-H)  \mathbf{v}^{2}
={r\sin^{2}\varphi}/{2\rho(r+\rho\cos\varphi)^{3}},\label{kvmHtorus}\end{equation}
and
\begin{equation}\nabla\cdot(\mathbf{m}\nabla\times\mathbf{m}) +\nabla\times(
\mathbf{m}\nabla\cdot\mathbf{m})=-\cos\varphi/\rho(
r+\rho\cos\varphi).\label{ntndmptorus}
\end{equation}

Additionally, from $\nabla^2(2H)dA=d\ast d(2H)$, we have
\begin{equation}\nabla^2(2H)=r(r\cos\varphi+\rho)/\rho^2(r+\rho\cos\varphi)^{3}.\label{nab2Htorus}\end{equation}

Substituting the above equations (\ref{HKtorus}) and
(\ref{nabvnabe3trs})--(\ref{nab2Htorus}) into Eq.\,(\ref{EL2}), we
can derive Eq.\,(\ref{toruseq}) in Sec.~\ref{Sectorus}.

\section{\label{dereqstwist}Derivation of equations (\ref{eqtwistr1})--(\ref{BC22}), and (\ref{fretwistr}) for twisted ribbons in Section \ref{SecTwR}}
A twisted ribbon can be expressed as a vector form
\begin{equation}
\mathbf{r}=\{u\cos \varphi ,u\sin \varphi ,\alpha \varphi
\}\label{vectortwist}
\end{equation}

A frame $\{\mathbf{e}_1,\mathbf{e}_2,\mathbf{e}_3\}$ is taken as
\begin{equation}\left\{\begin{array}{l}
\mathbf{e}_{1} =\{\cos \varphi ,\sin \varphi ,0\}, \\
\mathbf{e}_{2} =\{-u\sin \varphi ,u\cos
\varphi ,\alpha \}/\sqrt{u^{2}+\alpha ^{2}}, \\
\mathbf{e}_{3} =\{ \alpha \sin \varphi ,-\alpha \cos \varphi
,u\}/\sqrt{u^{2}+\alpha ^{2}}.
\end{array}\right.\label{framtwistr}\end{equation}

By using Eqs.\,(\ref{drframe}), (\ref{vectortwist}) and
(\ref{framtwistr}), we can derive
\begin{equation}
\omega _{1}=du,\omega _{2}=\sqrt{u^{2}+\alpha ^{2}}d\varphi.
\end{equation}
From Eq.\,(\ref{diffei}) and (\ref{framtwistr}), we obtain
\begin{eqnarray}
\omega _{12} &=&[u/(u^{2}+\alpha ^{2})]\omega _{2},\label{omega12twst} \\
\omega _{13} &=&-[\alpha /(u^{2}+\alpha ^{2})]\omega _{2},\label{omega13twst} \\
\omega _{23} &=&-[\alpha /(u^{2}+\alpha ^{2})]\omega
_{1}.\label{omega23twst}
\end{eqnarray}
Considering Eqs.\,(\ref{Sandomega12}), (\ref{nabdS}) and
(\ref{omega12twst}), we have
\begin{eqnarray}
&&\mathbf{S} =-[u/(u^{2}+\alpha ^{2})]\mathbf{e}_{2}, \\
&&\nabla \cdot \mathbf{S} =0.
\end{eqnarray}
Additionally, (\ref{omega13}), (\ref{omega13twst}) and
(\ref{omega23twst}) give
\begin{equation}
a=c=0,b=-\alpha /(u^{2}+\alpha ^{2}).\label{abctwist}
\end{equation}
Thus by considering Eqs.\,(\ref{torsionexp}), (\ref{HKgeneral}),
(\ref{normalcur}), and (\ref{laplphi}), we obtain
\begin{eqnarray}
&&H=0,K=-\alpha^{2}/(u^{2}+\alpha ^{2})^2,\label{HKtwist}\\
&&\kappa_\mathbf{m}=-\alpha\sin2\phi/(u^{2}+\alpha ^{2}),\label{kaptwistr}\\
&&\tau_\mathbf{m}=-\alpha\cos2\phi/(u^{2}+\alpha ^{2}),\label{tautwistr}\\
&&\nabla
^{2}\phi=\phi_{uu}+(u\phi_{u}+\phi_{\varphi\varphi})/(u^{2}+\alpha^{2}).
\end{eqnarray}
Substituting the above four equations into Eq.\,(\ref{EL3}), we can
derive Eq.\,(\ref{eqtwistr1}) in Sec.~\ref{SecTwR}.

If $\phi=0$ or $\pi/2$, then $\nabla\phi=0$, and
\begin{eqnarray}&&\mathbf{v}\equiv\nabla\phi-\mathbf{S} =[u/(u^{2}+\alpha ^{2})]\mathbf{e}_{2},\label{vecvtwst}\\
&&\mathbf{v}\cdot d\mathbf{e}_3=v_2\omega_{32}=[u\alpha
/(u^{2}+\alpha ^{2})^2] du,
\end{eqnarray}
where $v_2$ is the component of $\mathbf{v}$ in the direction of
$\mathbf{e}_2$. Thus from Eq.\,(\ref{nabvnabe3}) we can derive
\begin{equation}\nabla\mathbf{v}\colon\nabla\mathbf{e}_{3}=0.\label{nabvnabe3twst}
\end{equation}

Considering $\sin2\phi=0$ and $\cos2\phi=\pm 1$, from
Eqs.\,(\ref{kvmHgen}), (\ref{ntndmp}), (\ref{abctwist}) and
(\ref{vecvtwst}), we can obtain
\begin{eqnarray}&&(\kappa_{\mathbf{v}}-H)\mathbf{v}^2=0,\label{kvmHtwistr}\\
&&\nabla\cdot(\mathbf{m}\nabla\times\mathbf{m}) +\nabla\times(
\mathbf{m}\nabla\cdot\mathbf{m})=0.\label{ntndmptwst}
\end{eqnarray}

Substituting the above equations (\ref{HKtwist}) and
(\ref{nabvnabe3twst})--(\ref{ntndmptwst}) into Eq.\,(\ref{EL4}), we
can derive Eq.\,(\ref{eqtwistr2}) in Sec.~\ref{SecTwR}.

Now let us turn to the boundary conditions. If $\phi=0$ for
$\alpha<0$, we have $\bar{\phi}=\pm\frac{\pi}{2}$. At the boundary,
$\mathbf{t}\|\mathbf{e}_2$ and $u=u_0\equiv W/2$, so
\begin{equation}v_t=u_0/(u_0^{2}+\alpha^{2}),v_{b}=0.\label{vtvbtwst}\end{equation}
From Eqs.\,(\ref{HKtwist})--(\ref{tautwistr}), we can obtain
\begin{eqnarray}&&K=-\alpha^{2}/(u_0^{2}+\alpha^{2})^{2},\\
&&\kappa_{\mathbf{m}}=0,\tau_{\mathbf{m}}=-\alpha/(u_0^{2}+\alpha^{2}).\end{eqnarray}
Additionally, we have
\begin{eqnarray}
&&\hspace{-0.5cm}\kappa_n=a\cos^2(\pi/2)+b\sin\pi+c^2\sin(\pi/2)=0,\\
&&\hspace{-0.5cm}\kappa_g=\omega_{12}/\omega_2=u_0/(u_0^{2}+\alpha^{2}),\\
&&\hspace{-0.5cm}\tau_g=b\cos\pi+(c-a)(\sin\pi)/2=\alpha/(u_0^{2}+\alpha^{2}),\label{taubdtwist}
\end{eqnarray}
in terms of the definition of normal curvature, geodesic curvature
and geodesic torsion. From the above equations
(\ref{vtvbtwst})--(\ref{taubdtwist}), we find that among the
boundary conditions (\ref{BC1})--(\ref{BC4}), only Eq.\,(\ref{BC2})
is nontrivial, which gives Eq.\,(\ref{BC22}). Similarly, we obtain
the same equation as (\ref{BC22}) when $\phi=\pi/2$ for $\alpha>0$.

The average energy per area (\ref{fretwistr}) can be obtained from
$F/A$, where $A$ is the area of the twisted ribbon and $F$ is the
free energy (\ref{openFE}). The calculation is straightforward with
$c_0=0$, Eqs.\,(\ref{HKtwist}), (\ref{tautwistr}), (\ref{vecvtwst}),
and
\begin{eqnarray}
&&dA=\sqrt{u^{2}+\alpha ^{2}}dud\varphi,\\ &&ds=\sqrt{u_0^{2}+\alpha
^{2}}d\varphi.
\end{eqnarray}


\begin{thebibliography}{}
\bibitem{rmkspoint}The property of singular points on spherical
CLMs is still same as that on achiral spherical vesicles discussed
in \cite{VitelliPRE06}. The difference will be evident in
non-spherical CLMs.
\end{thebibliography}
\end{document}